\documentclass[onecolumn]{aastex62}

\usepackage{amsmath}
\usepackage{multirow}
\usepackage{makecell}

\def\Rsun{\textit{R}$_{\odot}$}
\def\Msun{\textit{M}$_{\odot}$}

\received{}
\revised{}
\accepted{}

\shorttitle{Mining black holes with LAMOST}
\shortauthors{Yi et al.}

\begin{document}

\title{Mining for Candidates of Galactic Stellar-mass Black Hole Binaries with LAMOST}

\correspondingauthor{Wei-Min Gu}
\email{guwm@xmu.edu.cn}

\author[0000-0002-5839-6744]{Tuan Yi}
\affil{Department of Astronomy, Xiamen University, Xiamen,
Fujian 361005, P. R. China}

\author[0000-0002-0771-2153]{Mouyuan Sun}
\affil{Department of Astronomy, Xiamen University, Xiamen,
Fujian 361005, P. R. China}

\author[0000-0003-3137-1851]{Wei-Min Gu}
\affil{Department of Astronomy, Xiamen University, Xiamen,
Fujian 361005, P. R. China}

\begin{abstract}
We study the prospects of searching for black hole (BH) binary systems
with a stellar-mass BH and a non-compact visible companion,
by utilizing the spectroscopic data of
\textit{Large Sky Area Multi-Object Fiber Spectroscopic Telescope} (LAMOST).
We simulate the Galactic BH binary population and determine its optical visibility
by considering the stellar synthetic population model and the distributions of binary orbital parameters.
By convolving the visibility of BH binaries with the LAMOST detection sensitivity,
we predict that $\gtrsim$ 400 candidate BH binaries can be found by the low-resolution, non-time-domain survey,
and $\sim$ 50-350 candidates by the LAMOST ongoing medium-resolution, time-domain spectroscopic survey.
Most of the candidates are short-period (0.2-2 days) binaries with M-, K-, G-, or F-type companions,
in which $\sim$ 47\% have a mass function (the lower limit of the BH mass) larger than 3 $M_{\odot}$.
By complementing the LAMOST spectroscopic data with other photometric/spectroscopic surveys
or follow-up observations, these candidates could be confirmed.
Therefore, by exploring the LAMOST data, we can enlarge the sample of
dynamically confirmed BH binaries significantly,
which can improve our understanding of the mass distribution of BHs and the stellar evolution model.
\end{abstract}



\section{Introduction} \label{sec:intro}

\renewcommand{\arraystretch}{1.5}
\setlength{\tabcolsep}{2.5pt}
\begin{center}
\begin{deluxetable*}{c|l|l}
\tablecaption{The distributions adopted in this paper. \label{table:dist}}
\tablehead{  \colhead{\textbf{Quantity}}  &  \colhead{\textbf{Distribution}}  &   \colhead{\textbf{Reference}} }
\startdata
\hline
 \multirow{2}{*}{ $M_{*}$ (\Msun) }
  & \textrm{Initial~Mass~Function~(IMF, subscript 0 means `initial'): }  &
\\
     &  \makecell{ \(\displaystyle \Psi_{M_{0}}(M_{*}) \propto
      \begin{cases}
      (M_{*}/0.5)^{-1.3} & \text{0.08 \Msun $\leqslant M_{*} < 0.5$~\Msun} \\
      (M_{*}/0.5)^{-2.3} & \text{0.5 \Msun $\leqslant M_{*} \leqslant 100$~\Msun}
      \end{cases}\)} & \citet{Kroupa2001}
\\
\hline
  \multirow{6}{*}{$M_{\rm BH}$ (\Msun)} & \textrm{BH~mass~distributions: fiducial model}   &
\\
   &  (a) $\Psi_{M_{\rm BH}}(M_{\rm BH})=\left\{A(M_{\rm BH})^n+
   \left[B(M_{\rm BH})^{-n}+C(M_{\rm BH})^{-n}\right]^{-1}\right\}^{1/n}$
   & \multirow{2}{*}{\citet{Ozel2010}}
\\
   & \textrm{where:~} $n=-10.0$,  $A(M_{\rm BH})=4.367-1.7294 M_{\rm BH}+0.1713 M_{\rm BH}^2$,
   &  \multirow{2}{*}{\citet{Ozel2012}}
\\
   & $B(M_{\rm BH})=14.24 \exp(-0.542 M_{\rm BH})$, $C(M_{\rm BH})=3.322 \exp(-0.386 M_{\rm BH})$   &
\\
  \cline{2-3}
   & \textrm{BH~mass~distributions: alternative model}  &
\\
   &  (b) $\Psi_{M_{\rm BH}}(M_{\rm BH})$ $\propto$ $ \exp(-k M_{\rm BH})$ &  \cite{Fryer2001}
\\
\hline
   \multirow{2}{*}{$a$ (\Rsun)} & \textrm{Logarithmically-flat~distributed~separation:}
   & \multirow{2}{*}{\citet{Abt1983} }
\\
   & $\Psi_{A}(a)$ $\propto$ $1/a$, ~3 \Rsun $\leqslant$ $a$ $\leqslant$ $10^{4}$ \Rsun &
\\
\hline
  \multirow{2}{*}{$i$}  & \textrm{Randomly-distributed~orbital~orientation:} &  \multirow{2}{*}{$-$}
\\
   & $\Psi_{I}(i) = {\rm Constant}, ~0 \leqslant i \leqslant \pi/2$  &
\\
\hline
     & \textrm{Galactic stellar~number~density~distribution: } &   \multirow{2}{*}{\citet{Juri2008}}
\\
   $n(r, \theta, \phi)$
   & $n(r, \theta, \phi)$ = $n_{0}[\exp(-\frac{r \sin \phi+R_{\rm Sun}}{2.6 {\rm kpc}}-\frac{r \cos \phi+Z_{\rm Sun}}{0.3 {\rm kpc}})$
   $+0.04 \exp(-\frac{r \sin \phi+R_{\rm Sun}}{3.6 {\rm kpc}})-\frac{r \cos \phi+Z_{\rm Sun}}{0.9 {\rm kpc}}]$
   & \multirow{2}{*}{\citet{Mashian2017}}
\\
   &  \textrm{where:~} $n_{0} \approx 3~{\rm pc}^{-3}$,~$R_{\rm Sun}=8~{\rm kpc}$,~$Z_{\rm Sun}=25~{\rm pc}$ &
\\
\enddata
\end{deluxetable*}
\end{center}

Stellar-mass black holes (BHs) are the ultimate fate of massive stars at the end of their life.
According to stellar evolution model,
there are around $10^8$-$10^9$ stellar-mass BHs reside in the Galaxy
\citep{Heuvel1992,Brown1994,Timmes1996,Agol2002}.
A significant amount of these dark objects are thought to exist in binary systems
(i.e., a system with a BH and a non-compact companion star).
So far, only about 60 of such systems were found \citep{Corral2016}.
Most of the candidates are interacting X-ray binaries \citep{McClintock2006,Remi2006},
i.e., the BH can accrete gas from its closely orbiting donor companion.

If the BH and its non-compact companion are detached \citep{Karpov2001,Yung2006}, namely,
they are still gravitationally bounded but the companion is not filling the Roche-lobe,
no mass-transfer-induced X-ray emissions will arise.
Examples of discovered isolated  and detached BH binaries are rare \citep{Casares2014N,Minniti2015,Giesers2018},
since these systems are hard to detect through the X-ray window.
Recently, the potential of detecting BHs by the mission of \textit{Gaia} was discussed intensively \citep{Breivik2017,Mashian2017,Yalinewich2018,Yamaguchi2018}.
\textit{Gaia} provides a dedicated way for hunting BHs by directly resolving the orbit astrometric signature \citep{Gaia2016}.
\cite{Mashian2017} estimated that $2 \times 10^5$ BHs could be hunted by \textit{Gaia},
while \citet{Breivik2017} predicted a roughly consistent result of 3800-12000 BHs by synthetic stellar evolution simulation.
Later, \cite{Yamaguchi2018} raised a key point that the effects of interstellar extinction ought to be considered,
which leads to a more moderate estimation of discovering 200-1000 BHs.
Another feasible way to identify short period candidates
is looking for photometric data that present characteristics
of microlensing and tidal distortion effects \citep{Wyrzy2016,Masuda2018}.
From the spectroscopic perspective, the system of a BH plus a non-compact companion is
a single-lined spectroscopic binary (SB1).
The stellar spectra provide information of stellar parameters
and radial velocities ($V_{\rm R}$) for the visible companion.
The mass function equation, derived from the radial velocity curve of the bright star,
puts a robust lower limit to the mass of the unseen star \citep[e.g.][]{Casares2014}.
If the orbital inclination is obtained through fitting the photometric light curves by synthetic models
\citep[e.g.][]{Orosz2000,Beer2002},
and the binary mass ratio constrained through, for instance,
resolving the rotational broadening of the photospheric lines from the visible companion \citep[e.g.][]{ Marsh1994},
the mass of the unseen object can be well determined.

The \textit{Large Sky Area Multi-Object Fiber Spectroscopic Telescope} \citep[LAMOST, ][]{Su2004,Cui2012}
is a unique astronomical instrument that has 4000 optical fibers assembled in the focal plane,
allowing the spectrographs to take thousands of spectra simultaneously.
The LAMOST Experiment for Galactic Understanding and Exploration
\citep[LEGUE;][]{Deng2012} survey is mobilized with
highly automated pipelines for sky subtraction \citep{Bai2017b}, cosmic ray removal \citep{Bai2017a},
data reduction and calibration \citep{Song2012,Luo2014}, and stellar parameters determination \citep{Wu2011,Wu2014,Xiang2015}.
Since its first data release in 2013 \citep{Luo2015},
the LAMOST spectroscopic survey has provided more than 10 million stellar spectra to date,
making it the largest stellar spectral database\footnote{\url{http://dr7.lamost.org/}} ever.
The ongoing LAMOST medium-resolution survey
has released data of around 3,124,000 stellar spectra with $S/N_{R}$ $>$10 at its first and second observing campaign.
The medium-resolution spectrographs have spectral resolution of $R \sim 7500$,
and the precision of the radial velocity measurements can reach 1 $\rm{km~s^{-1}}$ for $S/N_{R} \sim 20$ \citep{Liu2019}.
In the future, the ongoing LAMOST medium-resolution time-domain campaign will
provide high-quality multi-epoch spectroscopic observations,
i.e., each target will have more than 60 exposures.
Thus the LAMOST spectroscopic database is an
invaluable resource for mining BH binaries through the spectroscopic point of view,
i.e., with radial velocity measurements that provide dynamical constraints for binary systems.

In this paper, we focus on the prospects of mining candidate BH binaries with LAMOST.
This manuscript is organized as follows.
In Section \ref{sec:model}, we present our model.
In Section \ref{sec:results}, we show our results.
Summary and discussions are made in Section \ref{sec:conc}.


\section{Theoretical model} \label{sec:model}

In order to estimate the number of detectable BH candidates with LAMOST,
we must solve the following three questions: \\
(a) \emph{Population}:
    how many BH binary systems with stellar companions reside in our Galaxy? \\
(b) \emph{Observability}:
    how many of these BH binaries are within the detection limits of LAMOST? \\
(c) \emph{Identifiability}:
    how many of these BH binaries have a signature of being identified as candidates? \\
The first question concerns the population and the evolution paths of the BH binaries.
We start with some definitions that describe the orbital parameters of a binary system: \\
$M_{\rm min}$$=$$0.08$~\Msun: minimum mass of a hydrogen burning star. \\
$M_{\rm max}$$=$$100$~\Msun: maximal mass of a star that remains stable
under the balance of stellar gravity and its radiation pressure. \\
$M_1$ (\Msun): mass of the primary, the most massive component in the initially formed binary system. \\
$M_2$ (\Msun): mass of the companion, the least massive component in the initially formed binary system
($M_{\rm min}$ $\leqslant$ $M_2$ $\leqslant$ $M_1$ $\leqslant$ $M_{\rm max}$ are satisfied) . \\
$M_{\rm BH}$  (\Msun): mass of the BH as the remnant of a progenitor that ends its life. \\
$q$ $\equiv$ $M_2$/$M_{\rm BH}$: mass ratio of BH binary system.\\
$a$ (\Rsun): orbital semi-major axis (binary separation). \\
$P_{\rm orb}$ (days): orbital period of the binary.\\
$i$: orbital inclination angle.\\

\subsection{Population and Evolution Paths}
To estimate the BH binary population, we develop a similar methodology to
\citet{Mashian2017}.
It is widely believed that more than $\sim$ 50\% of stars reside in binary systems \citep{Duchene2013,Yuan2015}.
For massive early-type stars, the binary fraction even reaches $\sim$ 70\% in the open clusters \citep{Sana2012}.
BHs with stellar companions can be formed by different stellar evolution channels \citep{Belczynski2002,Heger2003}.
Typically, a massive progenitor with mass $\gtrsim$ 20\Msun~at its final evolution stage
will explode as a supernova (SN) and give birth to a remnant BH
by fallback accretion onto the collapsing core.
SN could lead to the destruction of the binary by the natal kick that
unbinds the remnant compact object and its companion \citep{Fryer2012}.
According to a recent study \citep{Koch2019}
on surviving star plus remnant binaries in a sample of 49 SN remnants,
$\lesssim$ 0.1 SN remnants contain a binary,
while the binary population synthesis models \citep{Belczynski2008} predict that
only $\sim$ 5\% of the star plus remnant binaries eventually survive the SN.
More massive progenitor with mass $\gtrsim$ 40\Msun~
may directly collapse into a BH without experiencing an SN explosion.
If we consider a \citet{Kroupa2001} initial mass function (IMF, Table~\ref{table:dist}, first row),
the fraction of  binaries that end up as a BH and a visible stellar companion is:
\begin{equation}\label{eq:frac}
f_{\rm BHB} = 0.5 \times (f_{\rm s} \int^{40 M_{\odot}}_{20 M_{\odot}} \Psi_{M_{0}}(M_{*}) dM_{*} +
\int^{100 M_{\odot}}_{40 M_{\odot}} \Psi_{M_{0}}(M_{*}) dM_{*})  \ ,
\end{equation}
where the factor 0.5 is assumed as the fraction of the binaries in the Galaxy,
namely, there are 50 binaries and 50 single stars in 150 stars.
$f_{\rm s}$ is the fraction of binaries that survived the SN explosion,
and $\Psi_{M_{0}}(M_{*})$ is the IMF with $M_{*}$ denoting the stellar mass. We assume $f_{\rm s}=0.05$.
The integration yields $f_{\rm BHB} \approx 0.0003$, i.e.,
roughly three BH binaries exist in every ten thousand stars.

\begin{figure*}
\figurenum{1}
\centering
\includegraphics[width=1.0\linewidth]{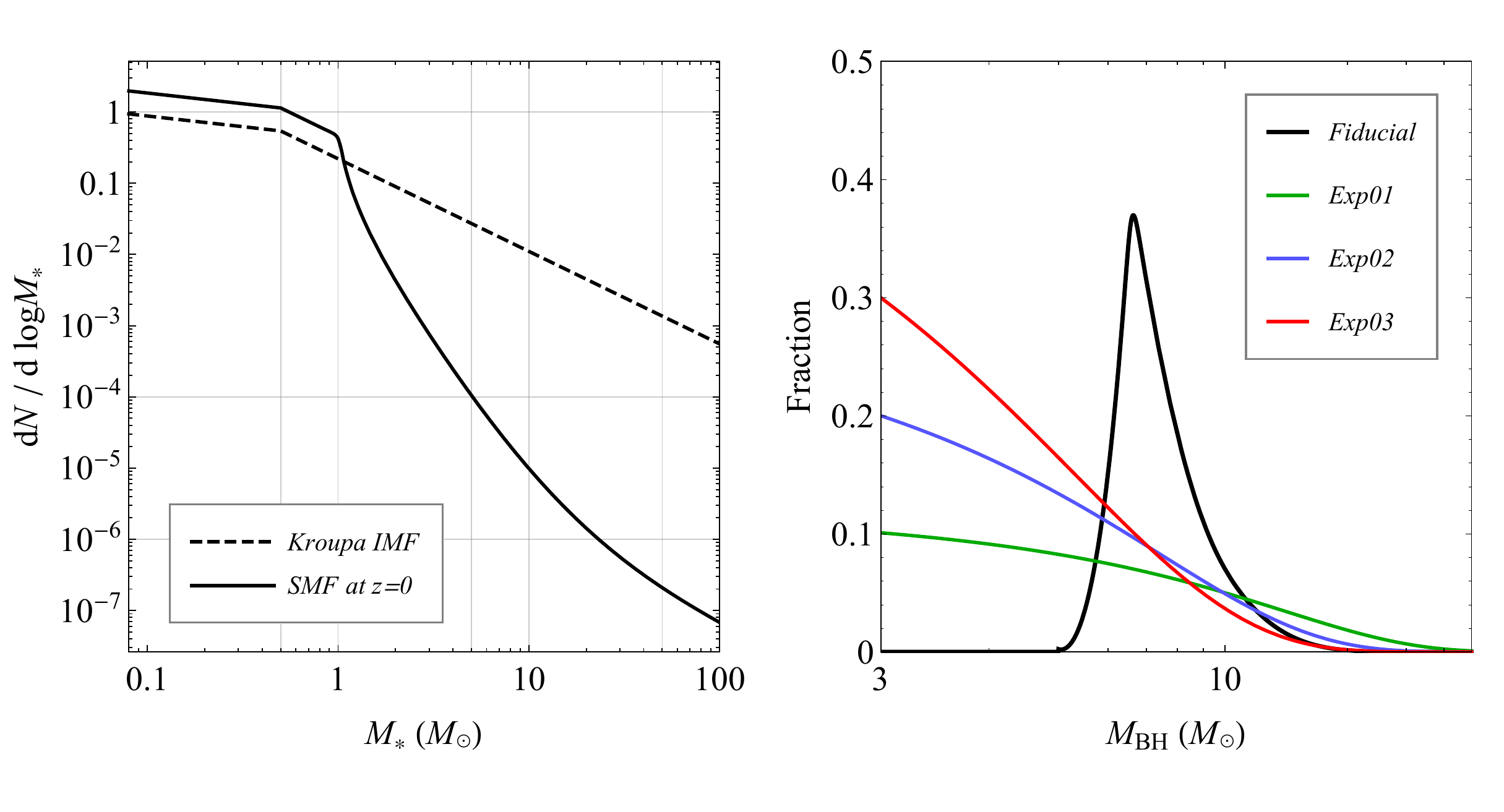}
\caption{Left panel: the IMF (dashed line) and the SMF (solid line).
Right panel: The fiducial BH mass distribution adopted from \citet[][black]{Ozel2010},
and the alternative exponential distributions motivated by \citet{Fryer2001}:
\textit{Exp01} (green), \textit{Exp02} (blue), and \textit{Exp03} (red)
correspond to the exponential factor $k$ =0.1, 0.2, and 0.3, respectively.
All distributions are normalized to make sure the integration over the mass range equals unity.
}
\label{F:1}
\end{figure*}

In order to detect the BH binaries via LAMOST,
the non-compact companion should be still shinning.
Following \citet{Mashian2017},
we calculate the fraction of stars that are still shinning as follows,
\begin{equation}\label{eq:fshinning}
f_{\rm shinning}=1-\frac{\rho_{*}(z(t_{\rm LB}=t_{\rm age}(M_{*})))}{\rho_{*}(0)}  \ ,
\end{equation}
where $M_{*}$, $\rho_{*}(z)$, $t_{\rm LB}$ and $t_{\rm age}(M_{*})$ are the stellar mass,
the comoving mass density\footnote{comoving mass density: the stellar mass density at redshift z.},
the look-back time\footnote{look-back time: the cosmological time at redshift z.},
and the age of the star, respectively.
Following \citet{Mashian2017}, we also integrate the star formation history \citep{Madau2014,Madau2017}
of the Universe to obtain $\rho_{*}(z)$;
the age of the star is estimated by using the analytical formula (Equation (4)) in \citet[][]{Hurley2000}.
Hence the stellar mass distribution (SMF) $\Psi_{M_2}(M_{*})$ of the visible companion at present ($z$=0) is:
\begin{equation}\label{eq:m2}
\Psi_{M_2}(M_{*}) \varpropto f_{\rm shinning}(M_{*})~\Psi_{M_{0}}(M_{*})\ .
\end{equation}
Shown in the left panel of Figure~\ref{F:1} are the IMF (dashed line) and the SMF (solid line).

The observability is also affected by the spatial distribution of BH binaries.
For simplicity, we assume that the spatial distribution of the binaries
traces the distribution of the stars in the Milky Way,
which can be modeled as a double exponential thick and thin disk model \citep{Gilmore1983,Juri2008}.
The adopted normalized number density profile is shown in Table~\ref{table:dist}
\citep[last row;][]{Mashian2017}.

It should be noted that during the evolution,
binary interactions \citep[e.g.][]{Hurley2002} can change the component masses and the orbital separation.
Processes such as mass loss by the stellar wind, mass transfer, mass accretion,
and common envelope (CE) evolution \citep[e.g.][]{Paczynski1976,Iben1993,Ivanova2013}
are not included in the current work for simplicity.
Binaries that have evolved through the CE phase
may have an orbital period (separation) orders of magnitude smaller than the initial period (separation),
since most of the angular momentum is taken away by the envelope \citep{Paczynski1976}.
We take a detour by adopting a logarithmically flat distribution for the orbital separation
\citep[Table~\ref{table:dist}, row 3;][]{Abt1983}.
The distribution can approximate the real cases, as it is a statistically summarized result from observations.

\subsection{Mass Distributions of BHs}
As for the mass distribution of BHs, we adopt the distribution of \citet{Ozel2010,Ozel2012}
as a fiducial one (hereafter referred to \textit{Fiducial} model).
\citet{Ozel2010} derived a BH mass distribution
from dynamical mass measurements of 16 BHs in transient low-mass X-ray binaries.
The distribution peaks at around 7 \Msun,
and presents a mass gap \citep[2-5 \Msun; see also][]{Bailyn1998,Farr2011,Belczynski2012}
between the population of BH and the most massive neutron star.
We also propose an alternative distribution for the purpose of comparison.
Our distribution is motivated by \citet{Fryer2001},
who studied the compact remnants masses by analyzing the balance
between the stellar bounding energy and supernova explosion energy.
\citet{Fryer2001} found that the BH mass distribution falls off exponentially.
Thus, we assume the BH mass distribution takes the form:
\begin{equation}\label{eq:dist01}
 \Psi_{M_{\rm BH}}(M_{\rm BH}) \propto \exp(-k M_{\rm BH}) \,
\end{equation}
where the exponential factor $k$ is taken to be 0.1, 0.2, 0.3 in the calculation.
Both distributions are shown in Table~\ref{table:dist} (second row) and plotted in the right panel of Figure~\ref{F:1}.

\subsection{Observational Cuts}
The low-resolution spectrographs of LAMOST are capable of observing stars
with an average limiting magnitude $m_{\rm V}^{\rm L}$ $\sim$ 18 mag in $V$-band and
a precision $V_{\rm R}^{\rm pre}$ =5 $\rm{km~s^{-1}}$ for radial velocity ($V_{\rm R}$) measurements \citep{Deng2012}.
The detection limits for medium-resolution spectrographs are:
$m_{\rm V}^{\rm L}$ $\sim$ 15 mag and $V_{\rm R}^{\rm pre}$ =1 $\rm{km~s^{-1}}$ \citep{Liu2019}.
The sky footprint of LAMOST is  $- \pi / 18$ $<$ $\delta$ $<$ $\pi / 3$ ($\delta$: the declination)
in the equatorial coordinate \citep{Zhao2012}.

To estimate the apparent brightness of binaries,
we assume the mass-luminosity relation $L=M_{\rm ZAMS}^{3.5}$
for zero-age main sequence (ZAMS) star.
Therefore, for a fixed distance $d$ (in units of pc),
there is a lower limit of the mass of the stars that are bright enough to be captured by LAMOST,
i.e., this lower mass limit is:
\begin{equation}\label{eq:mmin}
M_{\rm min}^{d} = \left( (\frac{d}{10~\rm{pc}})^{2} 10^{\frac{M_{\rm V}^{\rm Sun}-m_{\rm V}^{\rm L}+A_{\rm V}(d)}{2.5}} \right)^{\frac{1}{3.5}}\ ,
\end{equation}
where $M_{\rm V}^{\rm Sun}=4.83$ is the absolute magnitude of the Sun,
and $A_{\rm V}(d)$ is the interstellar extinction.
Following \citet{Yamaguchi2018}, we adopt the average extinction of the Galactic disk: 
$A_{\rm V}(d) = d/(1~{\rm kpc})$ \citep{Spitzer1978}.

There is also an accessible range of the radial velocity curve semi-amplitude $K_{2}$,
given the precision of $V_{\rm R}$ measurements.
From Kepler's third law, the semi-amplitude $K_{2}$ is:
\begin{equation}\label{eq:k2}
K_{2} = \left( \frac{G M_{\rm BH}}{a (1+q)} \right)  ^\frac{1}{2} \sin{i} \ .
\end{equation}
For the orbital inclination $i$, we assume that the binary orbits are randomly distributed (Table \ref{table:dist}, row 4).
The circular orbit assumption (zero eccentricity) is also assumed.

Let us define $\Psi_{K_{2}}(k)$ as the probability distribution function (PDF)
and $\Phi_{K_{2}}(k)$ as the cumulative distribution function (CDF) of $K_{2}$.
By definition,
\begin{equation}\label{eq:pdfk2}
\begin{aligned}
\Phi_{K_{2}}(k) = & \textit{Prob}(K_{2} \leqslant k) &
= \int_{K_{2} \leqslant k} \Psi_{A}(a) \Psi_{M_{\rm BH}}(m) \Psi_{Q}(q) \Psi_{I}(i) {\rm d}a{\rm d}m{\rm d}q{\rm d}i & \\
\Psi_{K_{2}}(k) = & \frac{\rm d}{{\rm d} k} \Phi_{K_{2}}(k) & &
\end{aligned} \ ,
\end{equation}
where $\textit{Prob}(K_{2} \leqslant k)$ is the probability of a specific binary system
whose amplitude $K_{2}$ is less than or equal to a measured value $k$.

We assure the reliability of $K_{2}$ measurements
by putting a lower cut $K_{2}^{\rm min}$ of 10 times $V_{\rm R}^{\rm pre}$.
In other words, the reliable measured $K_{2}$ should be $>$ 50 $\rm{km~s^{-1}}$ for low-resolution spectrograph,
and $>$ 10 $\rm{km~s^{-1}}$ for medium-resolution spectrograph.
So to acquire a proper error estimation of the amplitude $K_{2}$, i.e.,
for a radial velocity curve with semi-amplitude $K_{2}$ $>$ $K_{2}^{\rm min}$ = 50 $\rm{km~s^{-1}}$
taken by the low-resolution spectrograph, the uncertainty of $K_{2}$ is $<$ 5 $\rm{km~s^{-1}}$
provided that the average $V_{\rm R}$ uncertainty = 5 $\rm{km~s^{-1}}$.

Table~\ref{tab:cuts} summarizes the observational cuts for the LAMOST surveys.

\begin{table}
  \centering
  \caption{
	Observational cuts}
  \begin{tabular}{ccccc}
  \hline
  \multicolumn{1}{c}{survey} &
  \multicolumn{1}{c}{$R$} &
  \multicolumn{1}{c}{$m_{\rm V}^{\rm L}$ $^{(*)}$} &
  \multicolumn{1}{c}{ $K_{2}^{\rm min}$} &
  \multicolumn{1}{c}{$\delta$}
  \\
     & & ($\rm{mag}$) & ($\rm{km~s^{-1}}$) & ($\rm{radians}$)
  \\
  (1) & (2) & (3) & (4) & (5)
  \\
  \hline
  low-resolution        &  $\sim$ 1800   &   18   &  50 & -$\frac{\pi}{18}$ $\sim$ $\frac{\pi}{3}$  \\

  medium-resolution &  $\sim$ 7500   &   15   &  10 & -$\frac{\pi}{18}$ $\sim$ $\frac{\pi}{3}$  \\
  \hline
   \end{tabular}
\label{tab:cuts}
\tablecomments{Column (1): the LAMOST surveys.
 Column (2): the resolution of the spectrographs.  Column (3): the detection limit.
 Column (4): the detection limit of the $V_{\rm R}$ semi-amplitude.
 Column (5): the equatorial latitude.
 (*): The LAMOST low resolution survey can capture spectra
for stars brighter than $r \lesssim$  19 ($r$-band) during dark/grey time,
and $r \lesssim$ 17 or $J \lesssim$ 16 on nights that are moonlit or have low transparency \citep{Deng2012}.
}
\end{table}

\subsection{The Total Number of Candidates}
Based on the knowledge discussed above,
the number of detectable BH binary candidates can be estimated by a multi-dimensional integral:
\begin{equation}\label{eq:num}
\begin{aligned}
N = f_{\rm BHB}
   \times \int^{100 M_{\odot}}_{{\rm Min} [100, M_{\rm min}^{d}]} f_{\rm shinning}(M_{*})~\Psi_{M_{0}}(M_{*}) dM_{*} \\
   \times \underset{- \pi / 18 < \delta < \pi / 3}{\int \int}
             \sin \phi d\phi d\theta \int_{0}^{d} r^{2} dr n(r, \theta, \phi)
   \times \int^{\infty}_{10 V_{\rm R}^{\rm pre}} \Psi_{K_{2}}(k) {\rm d}k  \times f_{\rm cad}
\end{aligned} \ ,
\end{equation}
where the coefficient $f_{\rm BHB}$ is introduced in Equation~(\ref{eq:frac}).
The first integral constrains the fraction of visible stars that do not exceed LAMOST's detection threshold,
the integrand is introduced at Equation~(\ref{eq:m2}),
which calculates the fraction of stars that are still shining today by evolving the stellar population.
The lower limit of the first integral is calculated by using Equation~(\ref{eq:mmin}).
The second, third, and fourth integrals sum up the stellar number density in the Galactic coordinate system,
with the constraint of the equatorial declination:  $- \pi / 18$ $<$ $\delta$ $<$ $\pi / 3$.
The transformation of this LAMOST visible sky region from equatorial to Galactic coordinate system
is calculated by using Equations (1)-(4) in \citet{Poleski2013}.
The fifth integral is the fraction of the accessible range of $K_{2}$.
The last term $f_{\rm cad}$ in the equation is defined as
the fraction of sources with no less than three observations (spectra).
In our opinion, three observations is a necessary condition to set constraints on $K_{2}$. 

\subsection{Discovering and Confirming a BH}
So far we have covered the first (population) and the second (observability)
motivating questions given at the beginning of Section~\ref{sec:model}.
Now, discovering a potential candidate BH is one thing, confirming (identifying) is another.
Strategies are required to search for candidate BHs from large spectroscopic surveys.
Notably, \citet{Gu2019} adopted the relation of the stellar radius 
and the Roche-lobe radius to constrain the binary separations,
and used a few ($\geqslant$3) $V_{\rm R}$ measurements to constrain the lower limit of $V_{\rm R}$ excursions.
Their method was applied to search for BH candidates with LAMOST DR6.
\citet{Thompson2018} developed a strategy based on the maximum acceleration of the system,
to select potential candidates in SDSS APOGEE survey which has in average 2-4 measurements per system.
If one has sufficient $V_{\rm R}$ measurements, the radial velocity curve and hence the mass function can be obtained.
However, only a fraction of sources have intense observations in the LAMOST low-resolution survey.
For example,  $\sim 6\% $ of targets have three or more low-resolution visits in the LAMOST \textit{Kepler} field
\citep[spanning five years from 2012 to 2017;][Table1]{Zong2018}.
In this case, complementary follow-up spectroscopic and photometric observations
are required to measure the orbital periods.
For instance, \cite{Zheng2019} searched for BH candidates with orbital periods revealed by the ASAS-SN photometry.


\section{Results and analyses} \label{sec:results}

\begin{figure*}
\figurenum{2}
\centering
\includegraphics[width=1.0\linewidth]{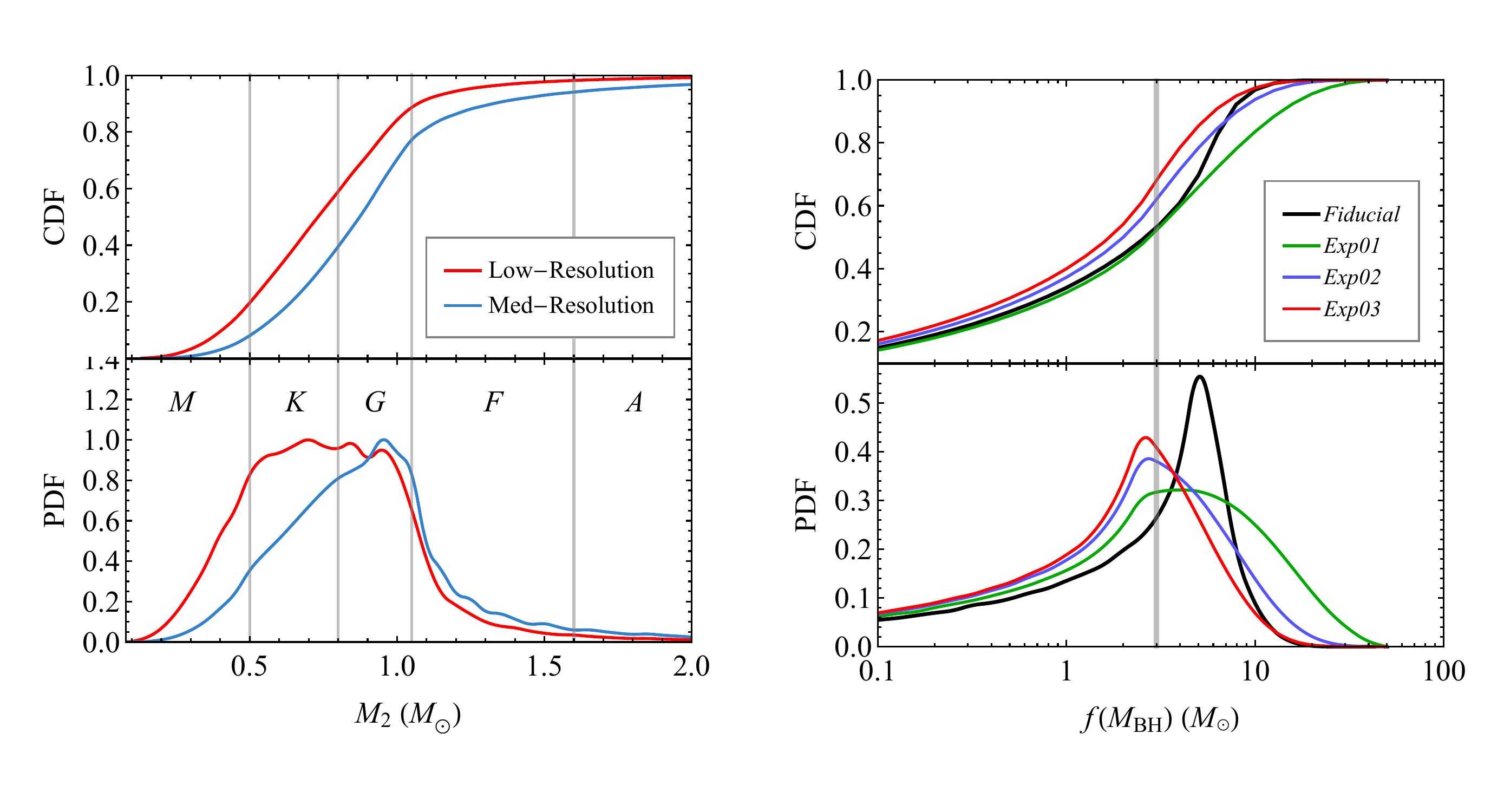}
\caption{The distribution (PDF) and cumulative distribution (CDF) of the visible companion's mass and mass function for the detectable BH binaries by LAMOST. Left panel: The PDF and CDF of
the visible companion's mass, for low- and medium-resolution spectrographs, respectively.
Most of the expected candidates have a visible companion of M-, K-, G-, or F-type star.
Right panel: The PDF and CDF of the BH binary mass function.
Around $47 \%$ of the BH binaries have a mass function larger than $3$~\Msun~(gray shallow line) for \textit{Fiducial} model.}
\label{F:2}
\end{figure*}

\begin{figure*}
\figurenum{3}
\centering
\includegraphics[width=0.67 \linewidth]{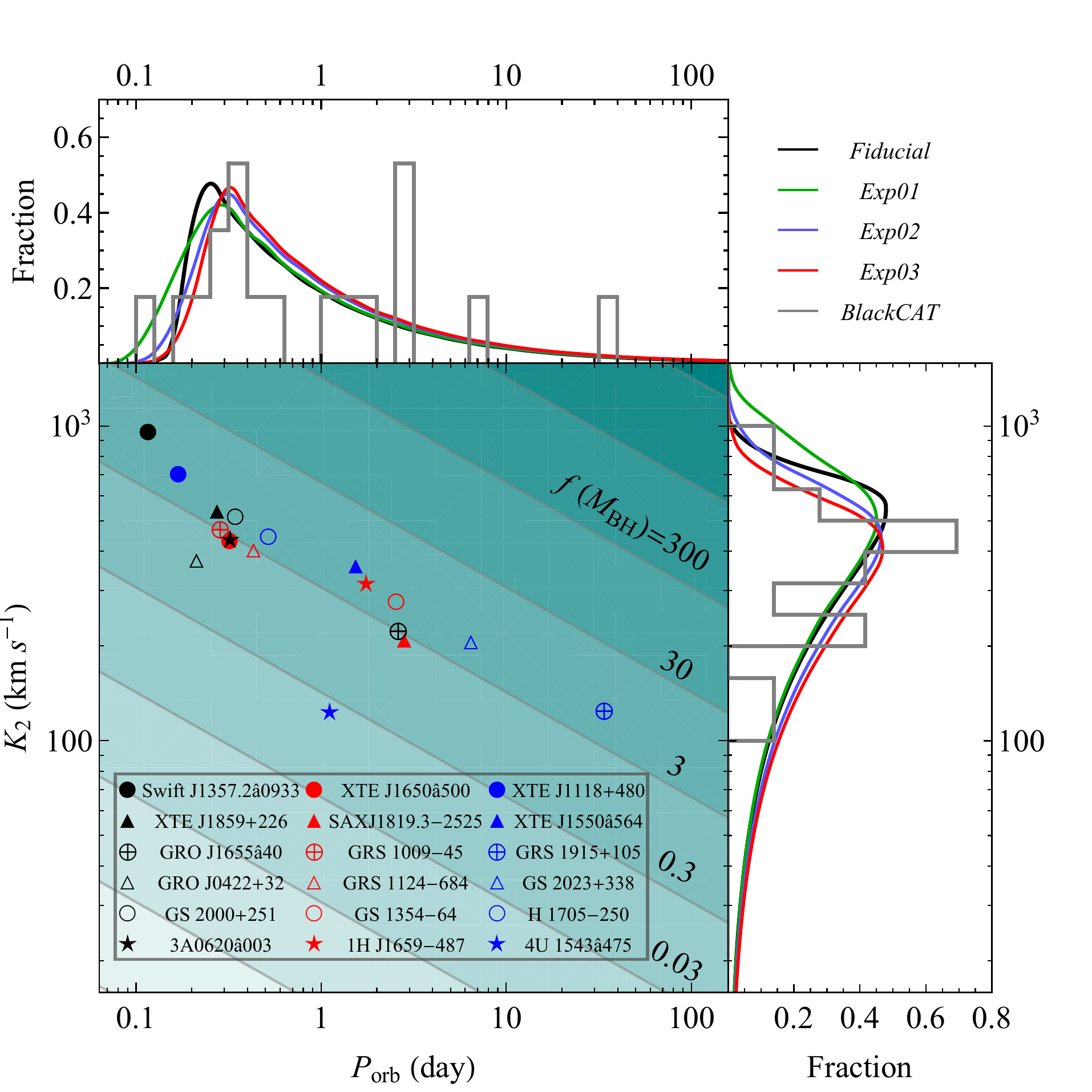}
\caption{The distributions of orbital period $P_{\rm orb}$
and radial velocity amplitude $K_{2}$ for visible companion.
Gray histograms denote the distributions of 18 dynamically confirmed BHs.
Data of $K_{2}$ and $P_{\rm orb}$ are adapted from BlackCAT
\citep[][Table A.4., columns 3 and 4]{Corral2016}.
The contours corresponding to $f(M_{\rm BH})$ = 0.03, 0.3, 3, 30, and 300 are also draw in the figure.
}
\label{F:3}
\end{figure*}

\begin{figure}
\figurenum{4}
\includegraphics[width=1.0\linewidth]{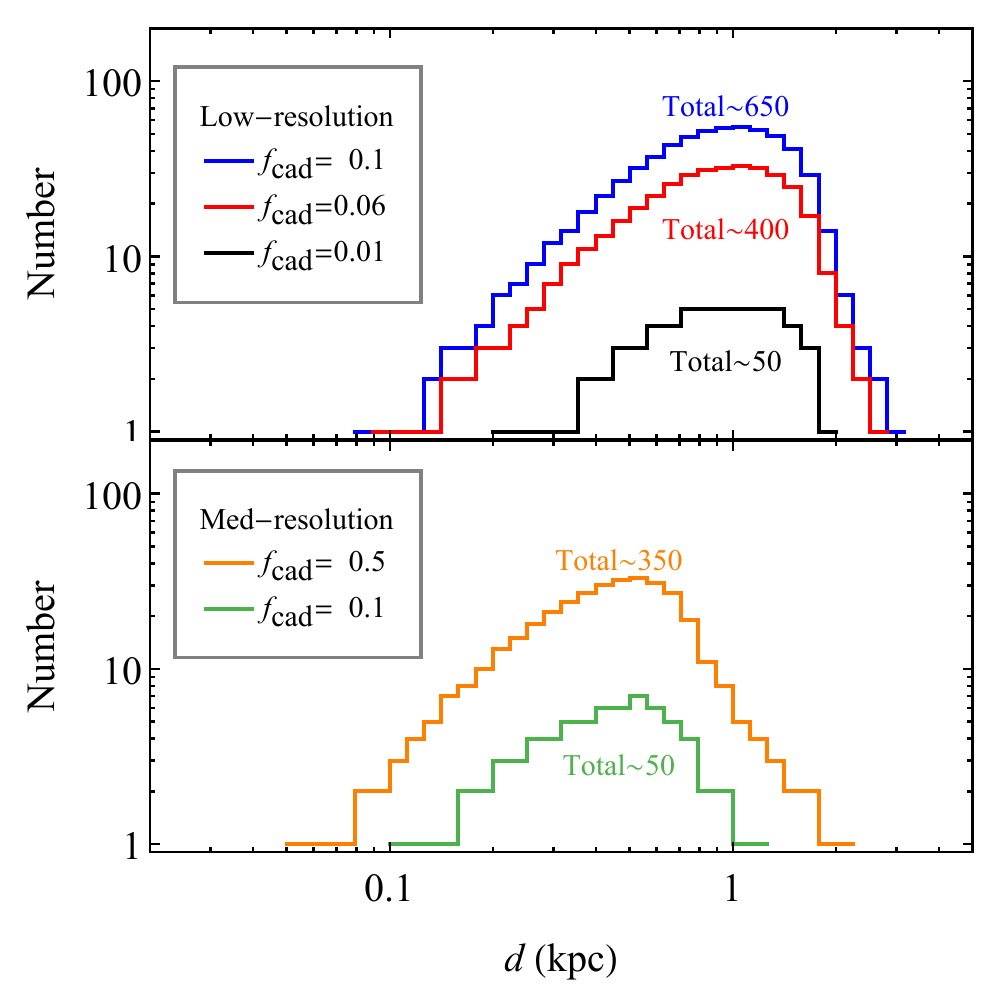}
\caption{The number of detectable candidates as a function of distance.
The blue, red and black lines represent the expected results by LAMOST
low-resolution, non-time-domain survey, with $f_{\rm cad}$ = $0.1$, $0.06$, and $0.01$, respectively.
The orange and green lines represent the expected results by medium-resolution, time-domain survey,
with $f_{\rm cad}$ = $0.5$, and $0.1$, respectively.
The total predicted numbers of each case are rounded to the nearest 50s.}
\label{F:4}
\end{figure}

Figure \ref{F:2} shows the distribution (PDF) and cumulative distribution (CDF)
of visible companion's mass and mass function for the detectable BH binaries by LAMOST.
The left panel is the PDF and CDF of the mass of detectable visible companions.
The distributions are calculated with equally spaced sampling points in the linear space of $M_{2}$,
with a bin size of 0.05 \Msun. By using Equation~(\ref{eq:num}),
we first calculate the number of BH binaries and the fraction of the observable companions
per mass bin at different distances, ranging from 20 \rm{pc} to 5 \rm{kpc},
then we sum up the number of each mass bin at all distances to find the fractions.
The red and blue lines represent the low- and medium-resolution spectrographs, respectively.
As suggested in the CDF, most of the detectable candidates are low-mass BH binaries with M-, K-, G- or F-type stars
\citep[MK system;][]{Morgan1973}, i.e., for low-resolution spectrographs,
the fractions for M-, K-, G- and F-type stars are $\sim$ 19.7\%, $\sim$ 39.3\%, $\sim$ 29.7\%, and $\sim$ 9.5\%, respectively.
The resulting fractions are a natural consequence of the relative number of stars with a given spectral type \citep{Chabrier2003},
plus the modification by the detection limits of LAMOST, i.e., fainter M-type stars are harder to be detected,
hence the fraction is quenched. As for massive early-type (O, B) stars,
the fraction of shinning ones is tiny because of the short life time scale of these populations.
As mentioned in Section \ref{sec:model}, however, binary fractions may be higher ($\sim$ 70\%) for early-type stars.
Consequently, $f_{\rm BHB}$ is evaluated to be $\sim$ 0.0004 in Equation~(\ref{eq:frac}),
suggesting that the population of BH plus early-type young star binaries may be underestimated.

As discussed in Section~\ref{sec:intro}, the binary mass function\footnote{Note the difference
between the initial mass function (IMF) and the binary mass function:
the IMF describes the mass distribution of the Galactic stellar population,
while the binary mass function is a measurable quantity for a specific SB1 system,
the lower mass limit of the unseen companion.}
is useful to constrain a lower mass limit of the unseen object in the SB1 system \citep{Casares2014}.
The binary mass function for a BH system is given by:
\begin{equation}\label{eq:massfun}
f(M_{\rm BH}) = \frac{M_{\rm BH} \sin ^{3} i}{(1+q)^{2}} = \frac{K^{3}_{2} P_{\rm orb}}{2 \pi G}\ .
\end{equation}
The right-hand side of the equation hints that
$f(M_{\rm BH})$ can be calculated by $K_{2}$ and $P_{\rm orb}$,
which are measurable quantities from the radial velocity curve.

The right panel of Figure~\ref{F:2} shows the PDF and CDF of the mass function.
The distributions are calculated with equally spaced sampling points in the logarithmic space of $f(M_{\rm BH})$,
with a step size of 0.1~\Msun. Comparing to the results derived from \textit{Exp} model,
the one from \textit{Fiducial} model rises more rapidly when passing 3~\Msun,
peaks at around 5~\Msun, and falls off more rapidly with increasing mass.
The vertical shallow gray line shows the BH binary system with $f(M_{\rm BH})=3$~\Msun.
Note that BH binaries may have a $f(M_{\rm BH}) < 3$~\Msun.
For instance, the mass function $f(M_{\rm BH})$ of SAXJ1819.3-2525 is $2.7 \pm 0.1$~\Msun~
\citep[V4641 Sgr, ][]{Orosz2001,Mac2014}.
In a practical perspective, if a SB1 system has mass function significantly larger than the mass of the visible star,
the unseen star must be a compact object.
In the calculation, we find that $\sim$ 47\%, $\sim$ 48\%, $\sim$ 38\%, and $\sim$ 32\%
of the BH binaries have mass functions larger than $3$~\Msun,
for the \textit{Fiducial}, \textit{Exp01}, \textit{Exp02}, and \textit{Exp03} models, respectively.

The distributions of the spectroscopic observables $P_{\rm orb}$ and $K_{2}$
are predicted and compared to 18 dynamically confirmed BHs \citep{Corral2016}.
The distributions of the $K_{2}$ values are calculated by Equation~(\ref{eq:pdfk2}).
By implementing Kepler's third law,
the distribution of $P_{\rm orb}$ values, for the fiducial and exponential $M_{\rm BH}$ models,
is derived from the distribution of orbital separations $a$ (Table~\ref{table:dist}, row 3).
Figure~\ref{F:3} shows the smoothed PDFs of  $P_{\rm orb}$ and $K_{2}$.
Gray histogram denotes the corresponding distribution
(fractional counts in the logarithmic space with a uniform step size = 0.1) of  the 18 dynamically confirmed BHs.
Data of $K_{2}$ and $P_{\rm orb}$ are adapted from BlackCAT
\citep[][Table A.4., columns 3 and 4]{Corral2016}.
The contours corresponding to $f(M_{\rm BH})$ = 0.03, 0.3, 3, 30, and 300 are also draw in the figure.

The orbital period distributions peak at around 0.23 (\textit{Fiducial})-0.3 (\textit{Exp01}) days,
with $\sim$ 76\% (\textit{Exp01})-78\% (\textit{Fiducial}) binaries in the range 0.2-2 days,
indicating that short period BH candidates are quite common in the observational space of parameters.
This implies that a large fraction will be quiescent BH transients, i.e., interacting binaries with accretion disks.
Quiescent BH transients can be easily identified from the presence of broad emission lines, chiefly $H_{\alpha}$.
In some cases, such as Swift J1357-0922 \citep{Mata2015},
the companion star is overwhelmed by the accretion disk light and,
consequently, no radial velocity information can be obtained.
In these cases, surveys exploiting the detection and properties of the $H_{\alpha}$ line
\citep{Casares2015,Casares2018} can be very useful.

Regarding the radial velocity semi-amplitude $K_{2}$ distribution,
the results show a peak at around 500 - 600~$\rm{km~s^{-1}}$.
We also present the calculated fractions in linear space, as presented in Table~\ref{table:frack}.
It shows that the fraction of $K_{2}$ values under 400~$\rm{km~s^{-1}}$ is larger than the fraction above,
indicating that there is a good chance of detecting BH binaries with $K_{2} <$ 400~$\rm{km~s^{-1}}$
(even for $K_{2} < $ 200~$\rm{km~s^{-1}}$, the odds are still good).
Note that extremely large $K_{2}$ $\sim$1000~$\rm{km~s^{-1}}$ might be possible.
These are close pairs that contain a low-mass dwarf (M dwarf) star
and a BH with $M_{\rm BH} \gtrsim$ 30~\Msun.
The BHs could form through the direct collapse of  massive O stars or Wolf-Rayet Stars,
without experiencing a supernova explosion and thus barely losing any mass.

We calculate Equation~(\ref{eq:num}) by setting
$f_{\rm cad}$=$0.06$ (see Section~\ref{sec:model}) for the low-resolution survey;
$f_{\rm cad}$=$0.1$ is also considered, since as the survey lasts, more spectra will be accumulated for most of the sources;
and $f_{\rm cad}$=$0.01$, to including less optimistic estimation.
As for the medium-resolution time-domain survey, sufficient observational epochs are guaranteed
($\sim$ 60 visits for every single target), leading to a tentative value of $f_{\rm cad}$=1.
However, the value of $f_{\rm cad}$ is unknown, so we take  $f_{\rm cad}$ = 0.5 and 0.1 for illustrative purposes.
Shown in the Figure~\ref{F:4} is the expected number of detectable candidates as a function of distance.
The calculation step size is 0.05 in logarithmic space of distance $d$.
We present only the results derived from \textit{Fiducial} distribution
as \textit{Exp} one gives same results.
The blue, red, and black lines represent the results by LAMOST
low-resolution, non-time-domain survey, for $f_{\rm cad}$ = $0.1$, $0.06$, and $0.01$, respectively.
The total number of each case are rounded to the nearest 50~s,
shown on top / below the lines (with the same color to the lines).
The number rises with increasing distance in the first place,
as the number of stars increases with the increasing volume.
The maximum number of the expected candidates is located near 1~{\rm kpc}.
At larger distances, the telescope gradually loses faint stars because of the detection limit
plus the effect of the Galactic extinction. The number falls off beyond 1~\rm{kpc},
with the farthest candidates located at $\sim$ 3~\rm{kpc}.
We expect $\sim$ 400 candidates to be found by the low-resolution survey,
depending on the actual value of $f_{\rm cad}$.
The orange and green lines represent the expected results by the medium-resolution, time-domain survey,
for $f_{\rm cad}$ = $0.5$, and $0.1$, respectively.
Compared to the low-resolution survey that can go deeper,
the medium-resolution survey can only catch stars that are brighter than 15~{\rm mag} in the V-band,
hence most of the candidates are expected to be located within 2~{\rm kpc}.
We expect $\sim$ 50-350 candidates to be found as the medium-resolution time-domain survey lasts.
The results show that both the low-resolution and medium-resolution surveys
will enlarge the size of current dynamically confirmed catalog by an order of magnitude.

\begin{deluxetable}{c|c|c|c|c|c|c}
\tablecaption{The distributions of $K_{2}$ in linear space. \label{table:frack}}
\tablehead{\colhead{Model} & \multicolumn{6}{c}{Fractions of $K_{2}$ at different range ($\rm{km~s^{-1}}$)} }
\startdata
\hline
                          & $<$ 200  & 200-400 & 400-600 & 600-800 & 800-1000  & $>$1000 \\
\hline
\textit{Fiducial}   & 0.355 & 0.328 & 0.245 & 0.068 & 0.003 & 0            \\
\hline
\textit{Exp01}      & 0.327 & 0.302 & 0.218 & 0.102 & 0.040 & 0.011     \\
\hline
\textit{Exp02}      & 0.392 & 0.346 & 0.200 & 0.052 & 0.009 & 0.001     \\
\hline
\textit{Exp03}      & 0.428 & 0.368 & 0.175 & 0.027 & 0.002 & 0            \\
\hline
\enddata
\end{deluxetable}


\section{Conclusions and discussion} \label{sec:conc}

In this paper, we study the prospects of searching for binary systems
with a stellar-mass BH and a non-compact visible companion,
by utilizing the spectroscopic data of LAMOST.
Our results can be summarized as follows. \\
$\bullet$ Most of the expected candidates have a visible companion of M-, K-, G-, or F-type star. \\
$\bullet$ About 47 \% of these BH binaries have a mass function larger than 3 \Msun. \\
$\bullet$ A majority of candidates have an orbital period in between 0.2-2 days,
 suggesting that the discovery of short period BH binaries is favored. \\
$\bullet$ Most of the detectable BH candidates are located within 2 {\rm kpc} of the solar neighborhood. \\
$\bullet$ We predict that $\gtrsim$ 400 candidate BH binaries can be found by the low-resolution,
non-time-domain survey, and $\sim$ 50-350 candidates by the LAMOST
ongoing medium-resolution, time-domain spectroscopic survey.
The results show that both low- and medium-resolution surveys
are promising to enlarge the size of current dynamically confirmed catalog by an order of magnitude.

We would like to address a caveat for the theoretical analyses in this work.
As mentioned in Section~2.1, binary interactions such as CE evolution are neglected for simplicity.
However, the impact on results remains uncertain.
For example, the CE phase can cause uncertainty on the second term in the right-hand side of Equation~(\ref{eq:frac}).
More specifically, in systems with large mass ratios, 
binaries with a massive progenitor and a low-mass late-type companion may eventually merge,
since the low-mass companion cannot take the burden of the CE mass ejection.
These systems thus have no contribution to the population of interest.
Hence the CE phase can reduce the population of BH binaries and hence decrease the discovery rates.
Binary population synthesis is needed for obtaining more precise results.

Recently, several works \citep{Breivik2017,Mashian2017,Yalinewich2018,Yamaguchi2018}
have investigated the ability to dynamically detect BH binaries by \textit{Gaia} satellite.
The \textit{Gaia} astrometric measurements resolve the orbital motion of the visible star,
such robust observations allow one to solve the mass of unseen object simply by Kepler's third law.
Since \textit{Gaia} has a unique strategy of sky-scanning that is designed for optimizing the astrometric accuracy,
it is better at detecting long period binaries ($\gtrsim$ 30 days)  over short period ones.
At this point, LAMOST has the advantage of observing binary systems with a wider range of orbital periods.
For each target, LAMOST takes three (or more) exposures in a single observation,
and each exposure takes 10-30 minutes.
Each target will be covered $\sim$ 60 times during the course of the entire time-domain survey.
Hence LAMOST is capable of tracking short period binaries down to a few hours,
or hunting long period binaries up to years.

As mentioned in Section~\ref{sec:model},
the potential of mining BH binaries from spectroscopic observations
can be reinforced with follow-up spectroscopic and photometric observations.
In fact, dynamically confirmed BHs are studied through both their
spectroscopy and photometry \citep[e.g.][]{Orosz1997,McClintock2001,Wu2016}.
On one hand, spectra provide information for the visible star's radial velocities, 
spectral types, and possible signature from an accretion disk (broad $H_{\alpha}$ emission line);
on the other hand,  light curves from photometric measurements provide valuable information
for the orbital period, orbital inclination, and the mass ratio of the binary
(in the case of extreme mass ratio, $q$ can be best obtained from 
resolving the rotational broadening ($V \sin i$) of the donor star \citep[e.g.][]{Marsh1994}).
It implies a novel approach for hunting a BH via the optical point of view:
(a) Start from picking out SB1 sources with large radial velocity excursions,
inspect whether broad emission lines are present.
(b) Cross-match suspected sources with other spectroscopic and photometric surveys,
set constraints upon the orbital period or the binary separation.
(c) Calculate the mass function of the unseen companion
and select candidates that have a mass function $f(M_{\rm BH})$ $>$ 3~\Msun.
(d) Collect data from follow-up observations and measure the full set of orbital parameters
that shall confirm candidates and put final constraints on the BH masses.

\acknowledgments

We thank Hao-Tong Zhang, Zhong-Rui Bai, Wei-Kai Zong, Xuefei Chen, and Hailiang Chen for beneficial discussion,
and thank the anonymous referee for giving numerous detailed, helpful suggestions that improved the manuscript.
This work was supported by the National Natural Science Foundation of China
under grants 11573023, 11603022 and 11973002, and the Fundamental
Research Funds for Xiamen University under grants 20720190122 and 20720190115.
Guoshoujing Telescope (the Large Sky Area Multi-Object Fiber Spectroscopic Telescope,
LAMOST) is a National Major Scientific Project built by the Chinese Academy of Sciences.
Funding for the project has been provided by the National Development and Reform Commission.
LAMOST is operated and managed by the National Astronomical Observatories, Chinese Academy of Sciences.


\end{document}